\documentclass[a4paper]{jpconf}

\usepackage[dvips]{epsfig}
\usepackage{amsfonts}
\usepackage{amssymb}
\usepackage{amscd}
\usepackage{pstricks}
\usepackage{amstext}
\input xy
\xyoption{all}

\newtheorem{example}{Example(s)}[section]

\newcommand\acknowlegements{{\bf Acknowlegements}\ }
\newtheorem{definition}[example]{Definition}
\newtheorem{proposition}[example]{Proposition}

\newtheorem{expl}[example]{Example}

\def\hs{\hbox to 3mm{}}
\def\hhs{\hbox to 5cm{}}
\def\ss{\smallskip}
\def\ms{\smallskip}
\def\bs{\bigskip}

\def\adots{\mathinner{\mkern2mu\raise1pt\hbox{.}
\mkern3mu\raise4pt\hbox{.}\mkern1mu\raise7pt\hbox{.}}}

\def\up#1{\raise 1ex\hbox{\footnotesize#1}}

\def\mref#1{{\footnotesize ({\ref{#1}})}}
\def\scal#1#2{\langle #1 | #2 \rangle}
\def\ncp#1#2{#1\langle #2\rangle}

\def\ra{\rightarrow}

\def\A{\mathcal{A}}
\def\B{\mathcal{B}}

\def\N{{\mathbb N}}
\def\C{{\mathbb C}}

\def\al{\alpha}

\def\ep{\epsilon}
\def\Si{{\Sigma}}

\begin{document}

\title{Hopf algebras: motivations and examples}

\author{G. H. E. Duchamp$^{\spadesuit}$, P. Blasiak$^{\diamondsuit}$, A. Horzela$^{\diamondsuit}$, K. A. Penson$^\heartsuit$ and A. I. Solomon$^{\clubsuit}{}^\heartsuit$}
\address{
$^\spadesuit$ Universit\'e Paris 13\\ Laboratoire d'Informatique Paris
Nord, CNRS UMR 7030\\99 Av. J-B. Cl\'ement, F 93430 Villetaneuse,
France\vspace{2mm} }
\address
{$^\diamondsuit$ H. Niewodnicza\'nski Institute of
Nuclear Physics, Polish Academy of Sciences\newline
ul.\ Eliasza-Radzikowskiego 152, PL 31342 Krak\'ow, Poland}
\address
{$^\heartsuit$ Laboratoire de Physique Th\'eorique de la Mati\`ere Condens\'ee\\
Universit\'e Pierre et Marie Curie, CNRS UMR 7600\\
Tour 24 - 2e \'et., 4 pl. Jussieu, F 75252 Paris Cedex 05, France}
\address{
$^\clubsuit$The Open University, Physics and Astronomy Department,
Milton Keynes MK7 6AA, United Kingdom}
\eads{\linebreak\mailto{ghed@lipn-univ.paris13.fr},
\mailto{penson@lptl.jussieu.fr} \linebreak}

\begin{abstract} This paper provides motivation as well as a method of construction for Hopf algebras, starting from an associative algebra. The dualization technique involved relies heavily on the use of Sweedler's dual.
\end{abstract}

\section{Introduction}

The formalism of quantum theory represents the physical concepts of states, observables
and their transformations as objects in some Hilbert space $\mathcal{H}$ (as we use a minimum of structure and instead of scalar products use duality, we will denote the space below by $V$). Briefly, vectors in the Hilbert space describe
states of a system, and linear transformations on $V$ represent basic observables.  Sets of transformations usually possess some structure such as that of a group, semi-group, Lie algebra, etc, and in general can be handled within the
concept of an algebra $\A$. The action of the algebra on the vector space of states $V$ and
observables $V^\vee$ is carried by its representation. Hence if an algebra is to describe physical
transformations it has to have representations in all physically relevant systems. This
requirement leads directly to the Hopf algebra structure in physics.
From the mathematical viewpoint the structure of the theory, modulo details, seems to
be clear. Physicists, however, need to have some additional properties and constructions
to move freely in this arena.\\ 
Hopf algebras have become ubiquitous in physics; we cite here only a few examples:
\begin{itemize}
	\item Feynman graphs are associated with an algebra related to the Hopf algebra of rooted trees \cite{CK1,CK2}. 
	\item In quantum statistical mechanics, the partition function description can be shown to give rise to a Hopf algebra structure – essentially a zero-dimensional analogue of the Feynman algebra cited above \cite{GOF5}. 
	\item Quantum groups, a special type of Hopf algebra, can provide useful models in quantum physics, for example giving a more accurate photon distribution for the laser than the usual Poisson one arising from the Glauber coherent states \cite{KS}.
\end{itemize}

Here we will show how the structure of Hopf algebras
enters  in the context of representations.

The first issue at point is the construction of the tensor product of vector spaces which is needed for the description of
composite systems. Suppose we know how a transformation acts on individual
systems, \emph{i.e.} we know its representation  in each of the vector spaces $V_1$ and
$V_2$, respectively. Hence the  need naturally arises for a canonical construction of an induced
representation of this algebra in $V_1\otimes V_2$ which would describe its action on the composite
system. Such a scheme exists and is provided by the co-product in the algebra, \emph{i.e.} a
morphism  $\Delta : \A\ra \A\otimes \A$. The physical plausibility of this construction requires
the equivalence of representations built on $(V_1\otimes V_2)\otimes V_3$ and $V_1\otimes (V_2 \otimes V_3)$, since
the composition of three systems cannot depend on the order in which it is done.
This requirement forces the co-product to be co-associative. Another point is connected
with the fact that from the physical point of view the vector space $\C$ represents a
trivial system having only one property ``being itself'', which cannot change.
Hence, one should have a canonical representation of the algebra on a trivial system,
denoted by $\ep : \A\ra \C$. Next, since the composition of any system with a trivial one
cannot introduce new representations, those on $V,\ V \otimes \C$ and $\C\otimes V$ should be equivalent.
This requirement imposes the condition on $\ep$ to be a co-unit in the (co-) algebra. In this
way we motivate the need for a bi-algebra structure in physics. The concept of an
antipode enters in the context of measurement. Measurement in a system is described
in terms of $V^\vee\otimes V$ and measurement predictions are given through the canonical
pairing $c : V^\vee\otimes V\ra C$. Observables, described in the dual space $V^\vee$, can also be
transformed and representations of appropriate algebras are given with the help of an
anti-morphism $S : \A \ra \A$. Physics requires that a transformation performed on the
system and apparatus simultaneously should not change the measurement results; hence
the pairing should trivially transform under the action of the bi-algebra. We thus obtain
the condition on $S$ to be an antipode, which is the last ingredient of a Hopf Algebra.

\section{How to never forget the requirements of Hopf algebras and give a meaning to them}

Let $\A$ be a $k$-associative algebra with unit (in short, a $k$-AAU; here $k$ will be a (commutative) field). We call a {\it representation} of $\A$ any mapping 
\begin{equation}
	\rho : \A\rightarrow End(V)
\end{equation}
which is compatible with the AAU structure of  both $\A$ and $End(V)$ (that is, $\rho$ is linear, $\rho(xy)=\rho(x)\circ\rho(y)$ identically\footnote{This is in order that $V$ be seen as a left module, see the ``additional discussion'' at the end of the manuscript.} and $\rho(1_\A)=Id_V$).

Usually, one considers the datum of $\rho$ as an ``action'' (on the left) of it over the $k$-vector space $V$ by
\begin{equation}\label{a_module}
	g.v:=\rho(g)(v)\ .
\end{equation}
This way of considering the action (given by Eq. \mref{a_module}) is just a change of notation and we will use either at times in order to give more flexibility or avoid confusion. Vector spaces with an action (as \mref{a_module}) of $\A$ on them are called (left) $\A$-modules.

\ss
One requires nice computation rules on $\A$-modules; that is, 
\begin{itemize}
	\item (direct) sums: given two $\A$-modules,  how do we construct a ``natural'' representation on $V_1\oplus V_2$
	\item products: same problem for $V_1\otimes V_2$, compatible with the associativity (\emph{i.e.} ``natural'' identification $V_1\otimes (V_2\otimes V_3)\simeq  (V_1\otimes V_2)\otimes V_3$ )
	\item unit: same problem for the ``natural'' identifications $V\otimes k\simeq k\otimes V\simeq V$ 
	\item a procedure to compare left and right modules; in particular, $V^\vee$ (the algebraic dual of $V$) possibly being compatible with the the natural pairings $V^\vee\otimes V\ra k$ and $V\otimes V^\vee\ra k$
\end{itemize}
 
(a) {\bf Sums}: If $\rho_i;\ i=1,2$ denote the representation morphisms on $V_i;\ i=1,2$. Then, one constructs
\begin{equation}
	\rho_3(u)=
	\left(\,\matrix{ 
	\rho_1(u) & 0_{Hom(V_1,V_2)}\cr
	0_{Hom(V_2,V_1)} & \rho_2(u) 
	}\,\right)
\end{equation}
where 
\begin{equation}
	a= 
	\left(\,\matrix{ 
	a_{11} & a_{12}\cr
	a_{21} & a_{22} 
	}\,\right)
\end{equation}
is the conventional representation of $a\in End(V_1\oplus V_2)$ subject to $a_{i,j}\in Hom(V_i,V_j);\ i,j=1,2$.

\ss
(b) {\bf Tensor products}: If one thinks about groups, with the notation as above, one has
\begin{equation}
	g.(x_1\otimes x_2)=g.x_1\otimes g.x_2
\end{equation}
($g$ acts as an automorphisms) and for Lie algebras $g$ acts as follows
\begin{equation}
	g.(x_1\otimes x_2)=g.x_1\otimes x_2+x_1\otimes g.x_2
\end{equation}
\emph{i.e.} by derivations. So, one can abstract the spaces and write that the scheme of action is $g\otimes g$ in case of groups and $g\otimes I+I\otimes g$ in the case of Lie algebras.\\
In either case (groups and Lie algebras) the representation can be rewritten as the action of an AAU $\A$ (algebra of the group in the first case and enveloping algebra in the second) and the ``scheme of action'' of $g\in \A$ can be written as a tensor of order two 
\begin{equation}
\sum g_{(1)}\otimes g_{(2)}\in \A\otimes \A.	
\end{equation}
In both these classical cases, one has
\begin{equation}
	g.(x_1\otimes x_2)=\sum g_{(1)}.x_1\otimes g_{(2)}.x_2\ .
\end{equation}
So, one needs a mapping $\Delta : \A\ra \A\otimes \A$, in Sweedler's notation \cite{Abe,Sw} 
\begin{equation}
\Delta(a)=\sum_{(1),(2)} a_{(1)}\otimes a_{(2)}	
\end{equation}
and one defines the action of $\A$ on a tensor product by 
\begin{equation}\label{tensor_rep}
	a.(x_1\otimes x_2)=\sum_{(1),(2)} a_{(1)}.x_1\otimes a_{(2)}.x_2\ .
\end{equation}
Now one can prove \cite{GOF13} that Eq. \mref{tensor_rep} defines a representation 
of $\A$ for any choice of $\rho_1,\ \rho_2$ iff $\Delta: \A\ra \A\otimes \A$ is a morphism of algebras \emph{i.e.} if, for all $u,v\in\A$ and with
\begin{equation}\label{uv_dec}
\Delta(u)=\sum_{(1),(2)}u_{(1)}\otimes u_{(2)};\ \Delta(v)=\sum_{(3),(4)}v_{(3)}\otimes v_{(4)}\ ,
\end{equation}
one has
\begin{equation}
	\Delta(uv)=\sum_{(1),(2),(3),(4)}u_{(1)}v_{(3)}\otimes u_{(2)}v_{(4)}.
\end{equation}
Moreover, one can prove that this procedure is compatible with the usual identification 
\begin{equation}
V_1\otimes (V_2\otimes V_3)\simeq  (V_1\otimes V_2)\otimes V_3
\end{equation}
for any choice of representations $\rho_1,\ \rho_2,\ \rho_3$ iff $\Delta$ is coassociative which means that  
\begin{equation}\label{co_ass}
(\Delta\otimes Id)\circ\Delta(u)=(Id\otimes \Delta)\circ\Delta\ .	
\end{equation}
The intuitive idea of the LHS of \mref{co_ass} applied to $u$ is obtained by splitting $u$ in two parts (as in eq \mref{uv_dec}) and then  resplitting the first component
$$
\Delta(u_{(1)})=\sum_{(11),(12)}u_{(11)}\otimes u_{(12)}
$$
then 
\begin{equation}\label{co_ass_left}
(\Delta\otimes Id)\circ\Delta(u)=\sum_{(11),(12),(2)}u_{(11)}\otimes u_{(12)}\otimes u_{(2)}
\end{equation}
and the RHS of \mref{co_ass} applied to $u$ is obtained by splitting $u$ in two parts and then resplitting the second component
$$
\Delta(u_{(2)})=\sum_{(21),(22)}u_{(21)}\otimes u_{(22)}\ .
$$
then 
\begin{equation}\label{co_ass_right}
(Id\otimes \Delta)\circ\Delta(u)=\sum_{(1),(21),(22)}u_{(1)}\otimes u_{(21)}\otimes u_{(22)}
\end{equation}
so, for all $u\in \A$, the result of the two computations \mref{co_ass_left}, \mref{co_ass_right} must coincide.

\ss
Another way to see co-asociativity which explains the name, and is also useful to produce examples and counterexamples, is through duality. For every $\phi,\psi\in \A^\vee$, one defines a linear form $\phi\ast \psi$ on $\A$ by 
\begin{equation}\label{dual_law}
	\scal{\phi\ast_\Delta \psi}{x}=\scal{\phi\otimes \psi}{\Delta(x)}\ .
\end{equation}
One can prove \cite{GOF13} that $\Delta$ is co-associative iff $\ast_\Delta$ is associative.

\ss
(c) {\bf Unit}: Here, one has to cope with the identifications $V\otimes k\simeq k\otimes V\simeq V$. One first needs a (one-dimensional) representation of $\A$ on $k$. This is exactly provided by a mapping $\ep:\A\ra k$ which is a morphism of AAU \emph{i.e.} a character. In order to have $V\otimes k\simeq V$ (resp. $k\otimes V\simeq V$), one needs to have identically (\emph{i.e.} for all $x\in \A$)
\begin{equation}\label{counit}
	\sum_{(1)(2)} x_{(1)}\ep(x_{(2)})=x; \ resp. \sum_{(1)(2)} \ep(x_{(1)})x_{(2)}=x\ .
\end{equation}
If $\ep$ fulfils these conditions one says that it is a counit. Again, one can invoke duality and transpose the character $^{t}\ep:k\ra \A^\vee$. Setting $e_\ep=^{t}\ep(1_k)$, one has that \cite{GOF13} $\ep$ is a counit iff $e_\ep$ is a unity in $(\A^\vee,\ast_\Delta)$, as defined in Eq. \mref{dual_law}. 

\ss
We summarize this discussion:
\begin{itemize}
	\item an AAU $\A$ is given
	\item there is no problem to compute direct sums of (same sided) representations of $\A$
	\item to construct tensor products one needs a mapping $\Delta: \A\ra \A\otimes \A$
	\item if one wants this construction to give representations of $\A$, one needs that $\Delta$ be a morphism of algebras
	\item if one requires that the identification $V_1\otimes (V_2\otimes V_3)\simeq  (V_1\otimes V_2)\otimes V_3$ be compatible with the construction of representations by $\Delta$, one needs that $\Delta$ be coassociative (\emph{i.e.} its dual be associative)
	\item for the identification $V\otimes k\simeq k\otimes V\simeq V$ to be compatible with the construction of representations by $\Delta$, one first needs a reprsentation of $\A$ on $k$, which is a character $\ep: \A\ra k$ and that $\ep$ be a counit (\emph{i.e.} its dual be a unit of the dual law $\ast_\Delta$).
\end{itemize}
At this point, if $(\A,\ast,1_\A,\Delta,\ep)$ fulfils the preceding requirements, $(\A,\ast,1_\A,\Delta,\ep)$ is called a bi-algebra.

\ss
The next paragraph will be devoted to two aspects of dualization: namely the dualization of the modules and of the algebra itself.

\section{Dualization, antipode and Sweedler's dual.}

\ss
The question of dualization is twofold.\\
Q1) Let $(\B,\ast,1_\A,\Delta,\ep)$ be a bialgebra. How does one  compare left and right $\B$-modules ? in particular, for a given module $V$, how to compare $V$ and $V^\vee$ ?\\
Q2) Can one endow $\B^\vee$, or at least a subspace of it, with the structure of a bialgebra by dualizing the product, coproduct, unit and counit ?

Throughout this paragraph $\A=(\A,\ast,1_\A,\Delta,\ep)$ is a bialgebra.

\subsection{Dualization of modules}

Let $V$ be a left $\A$-module, then $V^\vee$ is endowed with the structure of a right $\A$-module by 
\begin{equation}
	\scal{\psi.g}{x}=	\scal{\psi}{g.x}
\end{equation}
for $\phi\in V^\vee,\ g\in \A$ and $x\in V$. Then, how does one  compare $V^\vee$ and any left module $W$ ?\\ 
The solution is to reverse the multiplication of operators by use of an anti-endomorphism $S:\A\ra \A$, that is a linear mapping which is compatible with the identity $S(1_\A)=1_\A$ and reverses the products $S(xy)=S(y)S(x)$ (such an antimorphism should be familiar to the reader as the adjoint or transpose of a matrix). We then form a new action on the left of $V^\vee$ by 

\begin{equation}
	\scal{g\ast_S\psi}{x}=	\scal{\psi}{S(g).x}\ .
\end{equation}
It is left to the reader to check that this is a true action on the left.\\
If, moreover, one wants  this action to be compatible with the natural pairings $V^\vee\otimes V\ra k$ and $V\otimes V^\vee\ra k$, one should have, for every $\psi\in V^\vee,\ x\in V,\ g\in \A$ and $\Delta(g)=\sum_{(1)(2)}g_{(1)}\otimes g_{(2)}$
\begin{equation}
	\sum_{(1)(2)} \scal{g_{(1)}\ast_S\psi}{g_{(2)}x}=\sum_{(1)(2)} \scal{g_{(2)}\ast_S\psi}{g_{(1)}x}=\ep(g)\scal{\psi}{x}
\end{equation}
which is possible (in full generality) iff, for all $g\in \A$ 
\begin{equation}\label{antipode}
	\sum_{(1)(2)} S(g_{(1)})g_{(2)}=\sum_{(1)(2)} g_{(1)}S(g_{(2)})=\ep(g)1_\A\ .
\end{equation}
One can prove \cite{Abe,GOF13} that if there is a solution  $S$ of \mref{antipode}, it is unique and that it is an  antimorphism $\A\ra\A$. Such a solution, if it exists, is called an antipode for the bialgebra $\A$. Then, we arrive at the definition:

\begin{definition} An algebra $\A=(\A,\ast,1_\A,\Delta,\ep,S)$ is called a Hopf algebra iff $(\A,\ast,1_\A,\Delta,\ep)$ is a bialgebra and $S$ is an antipode for it.
\end{definition}
 
\begin{expl} Let $\Si=\{a,b,c,\cdots\}$ be an alphabet (\emph{i.e.} a set of variables) and $\Si^*$ be the free monoid i.e. the set of all words (or strings) in the letters of $\Si$. The set $\Si$ is endowed with the law of concatenation (which will be denoted ``$conc$'' \cite{Re} in this example) which is just the juxtaposition of the strings, namely with $u=a_1a_2\cdots a_p$ and $v=b_1b_2\cdots b_q$, one has
\begin{equation}
	conc(u,v)=uv=a_1a_2\cdots a_pb_1b_2\cdots b_q\ .
\end{equation}
This is an associative law with unit (the empty string $1_{\Si^*}$). The $k$ algebra of $\Si^*$ ($k[\Si^*]$) is usually denoted $\ncp{k}{\Si}$ and known under the name of algebra of non-commutative polynomials in the variables $\Si$ (or, due to its universal properties, the free algebra \cite{BR,B_Alg_III} and below). It is the set of linear combinations 
\begin{equation}\label{nocomm_poly}
	P=\sum_{i=1}^n \al_iw_i
\end{equation}
and, as in the commutative case, the $w_i\in \Si^*$ are called the monomials of $P$ and the $\al_i\in k$ its coefficients.
The product of two polynomials $P,Q$ as in \mref{nocomm_poly} is defined by the concatenation of their monomials (the strings $w_i$). It is easily checked that $(\ncp{k}{\Si},conc,1_{\Si^*})$ is an AAU \cite{BR,B_Alg_III} which is free in the following sense. For each set-theoretical mapping $\phi: \Si\ra\A$ where $\A$ is an AAU, there exists a unique morphism of AAUs $\bar\phi:\ncp{k}{\Si}\ra \A$ such that $\phi=\bar\phi\circ nat$

\begin{eqnarray}\label{free_alg}
\begin{array}{c}\xymatrix{
\ar[rr]^{\ \ \ \ \phi}\Si\ar[drr]_{nat \ \  \ \ \ \ }&&\A \\
 &&\ar[u]^{\bar\phi}\ncp{k}{\Si}\\
}
\end{array}
\end{eqnarray}
where $nat$ is the natural mapping which sends the letter $a\in \Si$ to itself considered as a monomial (of degree $1$). Given a partition of the alphabet, $\Si=G\sqcup L$ ($\Si$ is the union of the two disjoint subsets $G$ and $L$) and, for $x\in G$ (resp. $x\in L$), define 
\begin{equation}
\Delta(x)=x\otimes x\ ;\ \mathrm{(resp.}\ \Delta(x)=x\otimes 1 + 1\otimes x\mathrm{)}\ .
\end{equation}
One uses \mref{free_alg} to extend $\Delta$ as a mapping $\Delta: \ncp{k}{\Si}\ra \ncp{k}{\Si}\otimes \ncp{k}{\Si}$ which can be checked to admit the following nice combinatorial description. For a word $w=a_1\cdots a_n$ let us define $I_G$ (resp. $I_L$) as the set of places of letters in $G$ (resp. in $L$) 
\begin{equation}
	I_L=\{i\in [1..n]_\N\ |\ a_i\in L\}\ ;\ 	I_G=\{i\in [1..n]_\N\ |\ a_i\in G\}
\end{equation}
then 
\begin{equation}
	\Delta(w)=\sum_{I+J=I_L} w[I_G\cup I]\otimes w[I_G\cup J]
\end{equation}
where for $R=\{r_1,r_2\cdots r_k\}\subset [1..n]_\N$ (in increasing order), $w[R]$ is the subword $a_{r_1}a_{r_2}\cdots a_{r_k}$.\\
One can check that $(\ncp{k}{\Si},conc,1_{\Si^*},\Delta_{G,P},\ep)$ with $\ep(x)=1$ if $x\in G$ and $0$ if $x\in L$ ($\ep$ is extended to $\ncp{k}{\Si}$ using \mref{free_alg}).\\
The bialgebra $\ncp{k}{\Si}$ constructed above admits an antipode iff $G=\emptyset$.
\end{expl}

\subsection{Dualization of a bialgebra}

\bs
Given a  bialgebra $(\A,\ast,1_\A,\Delta,\ep)$, one wants to endow the dual space $\A^\vee$ with the structure of a bialgebra. Let us see, step by step what it is possible to do.\\
{\bf Coproduct and counit}: One can always transpose the comultiplication and the counit by the formulas given in eq. \mref{dual_law}, \mref{counit} and endow $\A^\vee$ with the structure of an AAU.\\
{\bf Dualization of the multiplication}: This is exactly the point where one has to restrict the domain because it is not always possible to dualize the multiplication operation on all elements of $\A^\vee$, since   
the transpose of the multiplication (denoted by  $\Delta_\ast$) is a mapping $\A^\vee\ra (\A\otimes\A)^\vee$ and, in general, its image is  not included in $\A^\vee\otimes\A^\vee$; the inclusion $(\A^\vee\otimes\A^\vee)\subset (\A\otimes\A)^\vee$ is strict in infinite dimension. Let us call $A^0$ the set of all elements $\psi\in A^\vee$ such that $\Delta_\ast(\psi)\in \A^\vee\otimes\A^\vee$. One may characterise this set  by use of the following finite orbit properties.

\begin{proposition}\label{finite_orbit} (see also \cite{Abe} paragraph 2.2, \cite{Sw} and \cite{GOF13})
Let $k$ be a field, $\A$ a $k$-algebra and $f\in \A^\vee$. For all $s\in \A$, we define the (shifted) linear forms $_sf,f_s$ by $_sf(x)=f(xs)$ and $f_s(x)=f(sx)$.\\ 
The following are equivalent : \\
i) The family $(f_s)_{s\in \A}$ is of finite rank in $\A^\vee$\\
ii) The family $(_sf)_{s\in \A}$ is of finite rank in $\A^\vee$\\
iii) There exists a double family $(g_i,h_i)_{1\leq i\leq n}$ of functions in $\A^\vee$ such that
\begin{equation}
\Big(\forall x,y\in S\Big)\Big(f(xy)=\sum_{i=1}^n\ g_i(x)h_i(y)\Big)
\end{equation}
iv) There exists $n\in \N$, $\lambda\in k^{1\times n},\gamma\in k^{n\times 1}$ and $\mu\ :\ \A\ra k^{n\times n}$ a morphism of algebras such that $(\forall s\in \A)(f(s)=\lambda\mu(s)\gamma)$. 
\end{proposition}

For proofs, see \cite{Abe,D21,DT,Ho,Ja,Sw}. The the set of elements which fulfil the equivalent conditions of proposition \mref{finite_orbit} is a subspace of $\A^\vee$ denoted $\A^0$, closed by all operations and co-operations, that is 
\begin{itemize}
	\item $\A^0$ is closed under $\ast_\Delta$, $\ep\in \A^0$ so that $(\A^0,\ast_\Delta,\ep)$ is a AAU
	\item $\Delta_\ast(\A^0)\subset \A^0\otimes \A^0$ and $\delta_{1_\A}: f\mapsto f(1_\A)$ is a counit for $\Delta_\ast$
	\item $(\A^0,\ast_\Delta,\ep,\Delta_\ast,\delta_{1_\A})$ is a bialgebra
	\item $^t S(\A^0)\subset \A^0$ and $^t S$ is an antipode for the bialgebra $(\A^0,\ast_\Delta,\ep,\Delta_\ast,\delta_{1_\A})$
\end{itemize}

A link with the theory of automata \cite{S1,S2} is that the elements of  Sweedler's dual of $\ncp{k}{\Si}$ are precisely the series that can be recognized by finite automata \cite{DT}.

\section{Conclusion}

In this note we introduced step by step the axioms of a Hopf algebra. The motivation for this is the following: An associative algebra with unit (AAU), which is a commonly occuring structure in physics, will inevitably lead to the necessity of dealing with representation modules, tensor products, duals, etc. The Hopf algebra structure provides us with just the machinery to manage these constructs. Further, the appropriate tools, such as product and co-product, occur in a symmetric fashion. In order to implement the passage to the symmetric, Hopf structure, a correct mathematical dualization requires the use of Sweedler's dual.

\ss\noindent
\acknowlegements The authors wish to acknowledge support from the Agence Nationale de la Recherche (Paris, France) under Program No. ANR-08-BLAN-0243-2 and from PAN/CNRS Project PICS No.4339(2008-2010). Two of us (P.B. and A.H.) wish to acknowledge support from the Polish Ministry of Science and Higher Education under Grants Nos. N202 061434 and N202 107 32/2832. Another (G.H.E.D.) would like to acknowledge support from ``Projet interne au LIPN 2009'' ``Polyz\^eta functions''.

\end{document}